\documentclass[9pt,twocolumn]{scrartcl}
\usepackage[dvipdfmx]{graphicx,color}
\usepackage{amsmath}
\usepackage{amssymb}
\usepackage{amsfonts}
\usepackage{multirow}
\usepackage{hhline}
\usepackage{url}

\title{A wide field-of-view crossed Dragone optical system using the anamorphic aspherical surfaces}

\author{Shingo Kashima$^1$, Masashi Hazumi$^{2,3,4,5}$, Hiroaki Imada$^{3,\ast}$, Nobuhiko Katayama$^4$, \\
Tomotake Matsumura$^5$, Yutaro Sekimoto$^3$, Hajime Sugai$^4$\\
\small{\textit{$^1$National Astronomical Observatory of Japan, Mitaka-shi, Tokyo, Japan.}}\\
\small{\textit{$^2$Institute of Particle and Nuclear Studies (IPNS), High Energy Accelerator Research Organization (KEK),}}\\
\small{\textit{Tsukuba, Ibaraki 305-0801, Japan.}}\\
\small{\textit{$^3$Institute of Space and Astronautical Science (ISAS), Japan Aerospace Exploration Agency (JAXA),}} \\
\small{\textit{Sagamihara, Kanagawa, 252-5210, Japan.}}\\
\small{\textit{$^4$Kavli Institute for the Physics and Mathematics of the Universe (IPMU),}} \\
\small{\textit{The University of Tokyo Institutes for Advanced Study, The University of Tokyo,}} \\
\small{\textit{Kashiwa, Chiba, Japan.}}\\
\small{\textit{$^5$The Graduate University for Advanced Studies (SOKENDAI),}}\\
\small{\textit{Miura District, Kanagawa 240-0115, Hayama, Japan.}}\\
\small{\textit{$^\ast$Corresponding author: imada@astro.isas.jaxa.jp}}
}

\begin{document}

\maketitle

\begin{abstract}
A side-fed crossed Dragone telescope provides a wide field-of-view. This type of a telescope is commonly employed in the measurement of cosmic microwave background (CMB) polarization, which requires an image-space telecentric telescope with a large focal plane over broadband coverage. We report the design of the wide field-of-view crossed Dragone optical system using the anamorphic aspherical surfaces with correction terms up to the 10th order. We achieved the Strehl ratio larger than 0.95 over $32\times18$ square degrees at 150~GHz. This design is an image-space telecentric and fully diffraction-limited system below 400 GHz. We discuss the optical performance in the uniformity of the axially symmetric point spread function and telecentricity over the field-of-view. We also address the analysis to evaluate the polarization properties, including the instrumental polarization, extinction rate, and polarization angle rotation. This work is a part of programs to design a compact multi-color wide field-of-view telescope for LiteBIRD, which is a next generation CMB polarization satellite.
\end{abstract}

\section{Introduction}
The measurements of the cosmic microwave background (CMB) have been providing the information on the physics of the early universe. While the temperature anisotropy of the CMB has been characterized in detail enough to establish the current modern cosmology, there is still an increasing interest of measuring the linear CMB polarization pattern called B-mode. The measurement of the primordial B-mode is sensitive to test the theory of inflation before the hot big bang and there is a world-wide competition to detect this signal\cite{planck2015,keckbicep2VI}.

The signal from the inflation is as low as nano Kelvin scale fluctuation at large angular scales (i.e., degree scale), compared with the uniform CMB temperature field at 3 Kelvin.  Thus, a corresponding telescope requires a capability to observe the full sky with high efficiency to increase the detection sensitivity. The goal of design optimization for such an optical system is often to maximize the telescope throughput, $ A \Omega $, where $ A$ is an entrance pupil area and $ \Omega $ is a solid angle corresponding to a field-of-view (FOV).

A crossed Dragone (CD) type, which consists of two mirrors of an offset paraboloid and an offset hyperboloid, is one of such survey-type telescopes \cite{tanaka1975,dragone1978}. The trade-off study by Tran et al. shows clear advantages of CD telescope configuration in the wide FOV, the telecentricity over a large focal plane area, and compactness in size~\cite{tran2008}. The side-fed CD configuration has been actually employed in CMB telescopes such as QUIET and ABS \cite{hanany2013, imbriale2011, essinger2011}, and the CD telescope also becomes a candidate of optical configuration for the post Planck CMB polarization satellite telescopes, including EPIC-IM and LiteBIRD \cite{tauber2014, tran2010, sugai2016}.

While the CD configuration provides a wide FOV without significant aberrations, there is a rapid increase of our interest to expand further its limit to populate an even larger number of detectors over the broader frequency coverage~\cite{matsumura2016}. Niemack has addressed the option to increase the available focal plane area by increasing the $F$ number, the ratio of focal length to effective aperture diameter~\cite{niemack2016}. This is an attractive approach when the optical envelope is not restricted, e.g. for a ground-based telescope. Also, a similar approach as we describe in this paper appears in de Bernardis et al.~\cite{arxiv1705.02170}.

We explore for the image-space telecentric and diffraction-limited system equipped with the wide focal plane without extending the optical envelope. This is particularly important for a balloon and satellite platform due to the limitation in space and mass. We aim at not only increasing the diffraction-limited area over broadband but also minimizing the asymmetry of the point-spread function (PSF) and the non-uniformity of the $F$ number over the focal plane. In this optimization, we employ anamorphic aspherical surfaces for the primary and secondary mirrors using the correction terms up to the 10th order. This allows to increase the degree-of-freedom in the design parameters with the presence of the multiple design constraints.

In this paper, we report the design to achieve the high Strehl ratio, the symmetry of PSF, and telecentricity over a wide FOV. We also evaluate the polarization properties, including the rotation of the electromagnetic plane, cross-polarization, and the instrumentally induced polarization. It is of our great interest to further address the optical properties using the physical optics simulation. The work is in progress and it is beyond the scope of this paper. This work is a part of the program to design a telescope for LiteBIRD, which is a next generation CMB polarization satellite.~\cite{ishino2016}. The detailed description of concept of the LiteBIRD optics is given by Sugai et al.~\cite{sugai2016}.

\section{Design and Optimization}
\label{dao}

\subsection{Presumptions}
\begin{figure}[t]
  \centering
   \includegraphics[width = 0.6 \hsize]{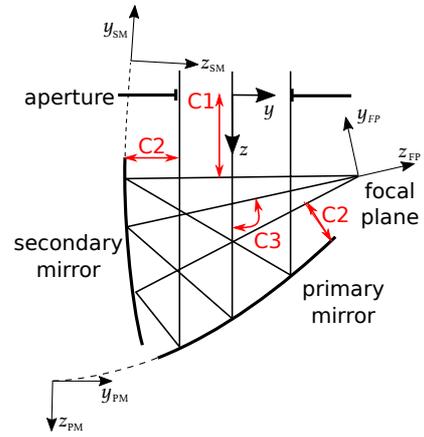}
   \caption{Schematic diagram of a CD telescope.The coordinates are set in order to define optical elements. C1, C2, and C3 represent the constraints used in optimization.}
   \label{fig:model}
\end{figure}

Figure~\ref{fig:model} shows a schematic view of a CD telescope that we optimize in this paper. It is assumed that LiteBIRD is equipped with this optical system and that a launch vehicle is an H2A or H3 rocket. The corresponding inner space which a telescope is allowed to use is a cylinder with a diameter of $\sim$\,1.6\,m and a height of $\sim$\,1.6\,m \cite{sugai2016}. This limitation to the whole size of a telescope restricts the $F$ number. An $ F $ number determines the space which the rays need to propagate, and then the mirror perimeters are determined not to vignette the rays. On the other hand, the mirrors can interfere each other mechanically when a too small F number is chosen. Thus, we consider only a CD telescope model with $ F \sim 3 $ and the entrance pupil diameter of 400~mm. The frequency range of our interest is 34-270~GHz.

\subsection{Design requirements}
We optimize a CD telescope to satisfy the following requirements over the FOV.
\begin{enumerate}
\item[R1.] Wave front aberration: $ > 0.95 $ in Strehl ratio (SR)
\item[R2.] The third flattening: $ < 0.05 $
\item[R3.] Telecentricity: Chief ray angle of incidence $ <1 $ degree
\item[R4.] Variations of $F$ number over the FOV: $ < 3 $\%.
\end{enumerate}

It is assumed that the vicinity of the PSF peak can be approximated to an elliptical Gaussian function. One of the characteristic parameter is the third flattening $ f $. It is defined as:
\begin{align}
f &:= \frac{ \sigma_\mathrm{ maj } - \sigma_\mathrm{ min } } { \sigma_\mathrm{ maj } + \sigma_\mathrm{ min } }, \\
I ( x_\mathrm{ FP }, y_\mathrm{ FP } ) & := I_0 \exp \left[ - 2 \left( \cfrac{ \left( x_\mathrm{ FP } - x_0 \right) \cos \theta
+ \left( y_\mathrm{ FP } - y_0 \right) \sin \theta } { \sigma_\mathrm{ maj } } \right)^2 \right. \nonumber \\
& \hspace{ 4 mm } \left. - 2 \left( \cfrac{ - \left( x_\mathrm{ FP } - x_0 \right) \sin \theta + \left( y_\mathrm{ FP } - y_0 \right) \cos
\theta } { \sigma_\mathrm{ min } } \right)^2 \right],
\label{eq:psfdef}
\end{align}
where $ \sigma_\mathrm{ maj } $ and $ \sigma_\mathrm{ min } $ are in major and minor axes, $ I_0 $ is the peak intensity, $ ( x_0, y_0 ) $ is the peak position on the focal plane, and $ \theta $ is the angle between $ x $ axis and the major axis.

The requirements are derived from the multiple drivers in the CMB experiment. The recent development of the multi-pixel array is based on the lithography technology, and thus a detector array is fabricated on a silicon wafer~\cite{matsumura2016}. Tiling such silicon wafers allows us to achieve a large focal plane detector array, and thus this drives the requirements, R1 and R3. Particularly, the detection efficiency becomes low when the beam does not enter the feed perpendicularly because the detected power is determined by the convolution of the field distributions from a telescope and a feed.

We introduce R4 to ensure that the coupling between the PSF and the beam pattern of the feed-horn is spatially uniform over the focal plane. If the target $ F $ number is set to 3 R4 means that the F number should be controlled within a range of 2.91 to 3.09. In this way, we do not have to fine-tune the feed-horn shape at the center and the edge of the plane.

Similarly, the far-field beam pattern is determined by this convolution. In general, a CMB experiment prefers working with an axially symmetric Gaussian beam for the simplicity in analysis and calibration. Thus, this drives the requirement R2. Note that the orientation of the ellipse of PSF is not restricted because only the magnitude of the third flattening can be a problem for a CMB observation. Although we did not include the polarization property as a design requirement, we assess its performance by evaluating the rotation of the electromagnetic plane, cross-polarization, and the instrumentally induced polarization in Sections~\ref{results}.\ref{sec:cx-pol} and \ref{results}.\ref{sec:ip}.

\subsection{Optimization}

We introduce the anamorphic aspherical surface as the following equation for both the primary mirror (PM) and the secondary mirror (SM):
\begin{align}
z_m & := \cfrac{ C_{m,x} { x_m }^2 + C_{m,y} { y_m }^2 } { 1 + \sqrt{ 1 - \left( 1 + k_{ m, x } \right) { C_{ m, x } }^2 { x_m }^2 - \left(
1 + k_{ m, y } \right) { C_{ m, y } }^2 { y_m }^2 } } \nonumber \\
& \hspace{ 5 mm } + \sum_{ i = 2 }^5 A_{ m, i } \left[ \left( 1 - B_{ m, i } \right) { x_m }^2 + \left( 1 + B_{ m, i } \right) { y_m }^2
\right]^i,
\label{eq01}
\end{align}
where $ m = \mathrm{ PM }, \mathrm{ SM }$, $ C_{ m, x } $ and $ C_{ m, y } $ are curvatures for $ x $ and $ y $ directions, $ k_{ m, x } $
and $ k_{ m, y } $ are conic constants in $ x $ and $ y $ directions, and $ A_{ m,i }$ and $ B_{ m, i } $ are aspherical coefficients. The
variables $ x_m $ and $ y_m $ are referenced to the local coordinates defining PM or SM, respectively (Fig. \ref{fig:model}). The origins of the coordinate $( x_m, y_m ) $ are located at $ ( x, y, z ) = ( 0, y_{ m, 0 }, z_{ m, 0 } ) $ and rotated about the $ x $ axis by an angle $ \theta_m $. The equation is an even function of $ x_m $ and $ y_m $, and the offsets to define mirrors are chosen in $y$ direction so that the symmetry at the $yz$ plane in Figure~\ref{fig:model} is maintained.

We conduct the optimization for three cases:
\begin{itemize}
  \item simple off-axis conic surface (SCS)
  \item anamorphic conic surfaces without higher-order terms (ACS)
  \item anamorphic aspherical surfaces with terms up to the 10$^{th}$ order (AAS).
\end{itemize}
The first case corresponds to the special case of Eq.~(\ref{eq01}) with $ C_{ m, x } = C_{ m, y } $, $ k_{ m, x } = k_{ m, y } $, and $ A_{
m, i } = B_{ m, i } = 0 $. The second case corresponds to $ C_{m,x} \neq C_{m,y} $ and $ k_{m,x} \neq k_{m,y} $, but $ A_{m,j} = B_{m,j} = 0$. The third case is Eq.~(\ref{eq01}) itself including all the parameters. 

The optimization is carried out with the following constraints.
\begin{enumerate}
  \item[C1.] Distance between aperture and rays: $ > 120$~mm,
  \item[C2.] Distance between rays and mirrors: $ > 30 $~mm,
  \item[C3.] Crossed angle of the optical axis between the primary and secondary mirrors: about 112 degs.
\end{enumerate}
These constraints are shown in Fig. \ref{fig:model}. The first constraint is motivated by keeping room to equip a baffling structure between aperture and rays, as well as by avoiding the mechanical interference between the aperture and the focal plane. The second one is to keep the electromagnetic waves reflected or diffracted at another optical element away from the mirror of interest. For example, we keep the electromagnetic wave diffracted at the aperture 30-mm away from the secondary. 30 mm corresponds to a few wavelengths in the lowest frequency. The third constraint is set to increase the mechanically available focal plane area and also to minimize stray light within this compact telescope configuration. We choose this value as a reasonable choice but the angles around 112 degrees still achieve the similar functionality.

 For the simplicity, the optimization is done only at 150~GHz because there is no chromatic aberration for a mirror system. We evaluate the other frequency ranges once the optimization is completed. The optimization and the analysis are done by using CodeV \cite{CodeV}. The SR quoted in this paper is associated with the wave front error via approximation employed by Mahajan \cite{mahajan1983}. In order to evaluate the variation of the $F$ number and the PSF over the FOV, we employ the ABCD matrix method \cite{gerrard2012, buchdahl1993}. To satisfy the requirement R2, the $ F $ numbers in $ x $ and $ y $ directions are controlled instead of the third flattening itself. After optimization, a PSF and third flattening are calculated and confirmed whether they satisfy the requirement R2.

\section{Results} \label{results}
\begin{figure}[t]
  \centering
   \includegraphics[width = 0.8 \hsize]{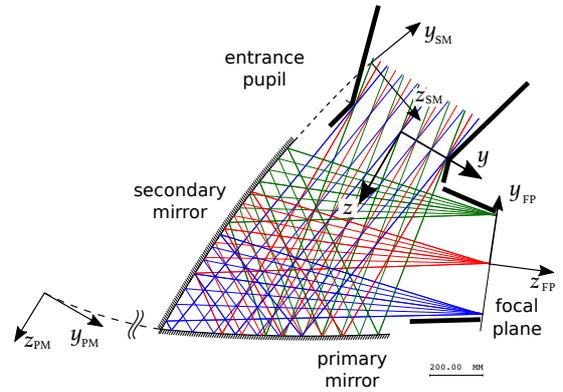}
   \caption{Cross-sectional view of the optical layout of the baseline design, which has the $F$ number of 3 and the entrance pupil diameter of 400~mm. The black thick lines demonstrate an example of a baffling structure.}
   \label{fig:layout}
\end{figure}

Fig. \ref{fig:layout} shows the layout obtained in our optimization for the AAS case. Tables \ref{tab_param_SCS}, \ref{tab_param_ACS}, and \ref{tab_param_AAS} show the derived parameters for the three cases, SCS, ACS, and AAS. The results and discussion in the present paper are based on these parameters.

\subsection{Mirror sag}
\begin{figure}[t]
   \centering
 \includegraphics[width=3.8in]{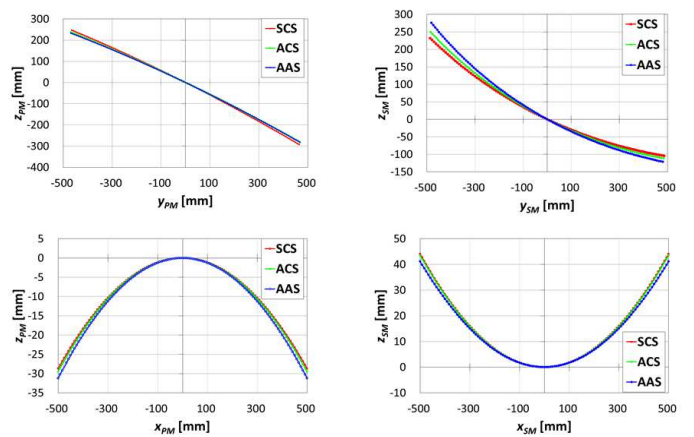}
    \caption{The primary and secondary mirror shapes are overplotted for three type of the surface corrections. The three surfaces are coincided at the center of the mirror.}
   \label{fig:surface_shapes}
\end{figure}

Fig.~\ref{fig:surface_shapes} shows the slice of the primary and secondary mirrors of the three surface types. There is no apparent higher order spatial variation due to the higher order corrections. When we assume the size of each surface to be 1~m and coincide the center of the mirror, the largest differential sag among the three types is 3~mm between the AAS and the SCS for the primary mirror and 12~mm for the secondary mirror.

\begin{table*}[t]
\centering
\caption{The parameters for SCS as the result of optimization.}
\label{tab_param_SCS}
\begin{tabular}{c|cc|ccc} \hline
 & $ C_m / \mathrm{ mm }^{ -1 } $ & $ k_m $ & $ y_{ m, 0 } / \mathrm{ mm } $ & $ z_{ m, 0 } / \mathrm{ mm } $ & $ \theta_m / \mathrm{ deg. } $ \\ \hline
PM & $ -2.2965 \times 10^{ -4 } $ & $ -0.8524 $ & $ -2616.51 $ & $ 1715.72 $ & $ 0 $ \\
SM & $ 3.6274 \times 10^{ -4 } $ & $ -3.3990 $ & $ -246.17 $ & $ -56.72 $ & $ 73.025 $ \\
FP & $ 0 $ & -- & $ 578.71 $ & $ 156.70 $ & $ 112.048 $ \\ \hline
\end{tabular}
%
\caption{The parameters for ACS as the result of optimization.}
\label{tab_param_ACS}
\begin{tabular}{c|cccc|ccc} \hline
 & $ C_{ m, x } / \mathrm{ mm }^{ -1 } $ & $ C_{ m, y } / \mathrm{ mm }^{ -1 } $ & $ k_{ m, x }$ & $ k_{ m, y } $ & $ y_{ m, 0 } / \mathrm{ mm } $ & $ z_{ m, 0 } / \mathrm{ mm } $ & $ \theta_m / \mathrm{ deg. } $ \\ \hline
PM & $ -2.2989 \times 10^{ -4 } $ & $ -2.2119 \times 10^{ -4 } $ & $ -0.4978 $ & $ -0.8094 $ & $ -2604.93 $ & $ 1610.07 $ & $ 0 $ \\
SM & $ 3.5203 \times 10^{ -4 } $ & $ 3.7695 \times 10^{ -4 } $ & $ -2.9848 $ & $ -3.3552 $ & $ -252.586 $ & $ -172.5887 $ & $ 71.342 $ \\
FP & $ 0 $ & $ 0 $ & -- & -- & $ 541.27 $ & $ 169.91 $ & $ 112.001 $ \\ \hline
\end{tabular}
%
\caption{The parameters for AAS as the result of optimization.}
\label{tab_param_AAS}
\begin{tabular}{c|cccc|ccc} \hline
 & $ C_{ m, x } / \mathrm{ mm }^{ -1 } $ & $ C_{ m, y } / \mathrm{ mm }^{ -1 } $ & $ k_{ m, x }$ & $ k_{ m, y } $ & $ y_{ m, 0 } / \mathrm{ mm } $ & $ z_{ m, 0 } / \mathrm{ mm } $ & $ \theta_m / \mathrm{ deg. } $ \\ \hline
PM & $ -2.4126 \times 10^{ -4 } $ & $ -2.1942 \times 10^{ -4 } $ & $ -0.6361 $ & $ -1.0535 $ & $ -2514.33 $ & $ 1588.89 $ & $ 0 $ \\
SM & $ 3.5710 \times 10^{ -4 } $ & $ 3.8091 \times 10^{ -4 } $ & $ -4.8176 $ & $ -3.5555 $ & $ -211.92 $ & $ -179.00 $ & $ 69.240 $ \\
FP & $ 0 $ & $ 0 $ & -- & -- & $ 543.02 $ & $ 230.75 $ & $ 111.955 $ \\ \hline
\end{tabular}
\begin{tabular}{c|cc|cc|cc|cc} \hline
& $ A_{ m, 2 }$ & $ B_{ m, 2 } $ & $ A_{ m, 3 }$ & $ B_{ m, 3 } $ & $ A_{ m, 4 }$ & $ B_{ m, 4 } $ & $ A_{ m, 5 }$ & $ B_{ m, 5 } $ \\ \hline
PM & $ -4.4252 \times 10^{ -12 } $ & $ -0.9958 $ & $ 1.3533 \times 10^{ -20 } $ & $ -0.6486 $ & $ -1.5402 \times 10^{ -29 } $ & $ 1.2958 $ & $ -1.9191 \times 10^{ -34 } $ & $ -0.06712 $ \\
SM & $ -8.8005 \times 10^{ -14 } $ & $ 0.04926 $ & $ 7.6442 \times 10^{ -19 } $ & $ 0.04683 $ & $ -1.0061 \times 10^{ -26 } $ & $ 0.8811 $ & $ -7.4764 \times 10^{ -31 } $ & $ -0.5148 $ \\ \hline
\end{tabular}
\end{table*}

\begin{figure}[!t]
  \centering
     \includegraphics[width=\hsize]{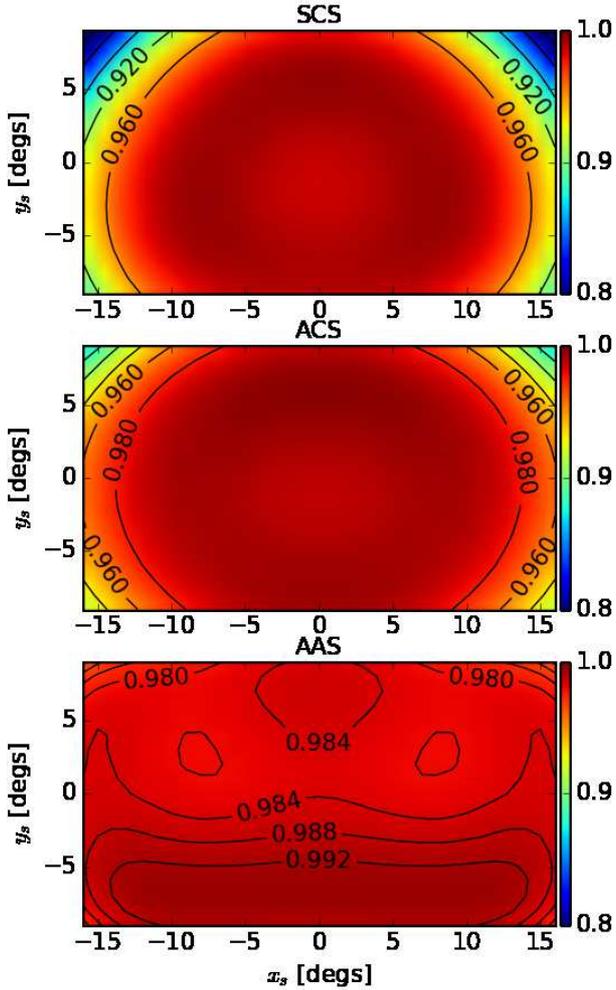} 
   \caption{The Strehl ratio of the optics for three different surface types evaluated at 150 GHz.}
   \label{fig:SR}
\end{figure}

\begin{figure}[!t]
   \centering
    \includegraphics[width=0.8\hsize]{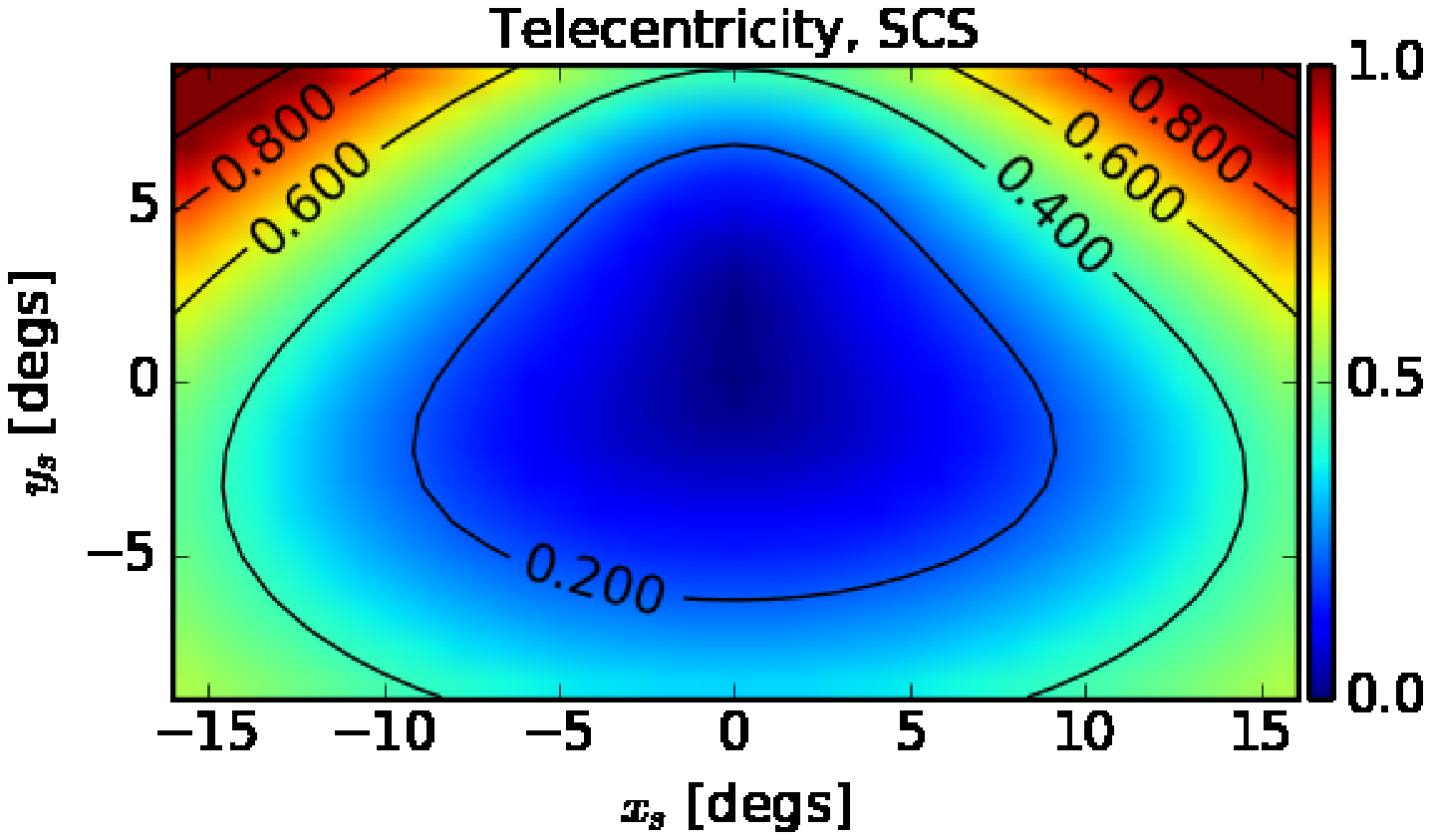}
    \includegraphics[width=0.8\hsize]{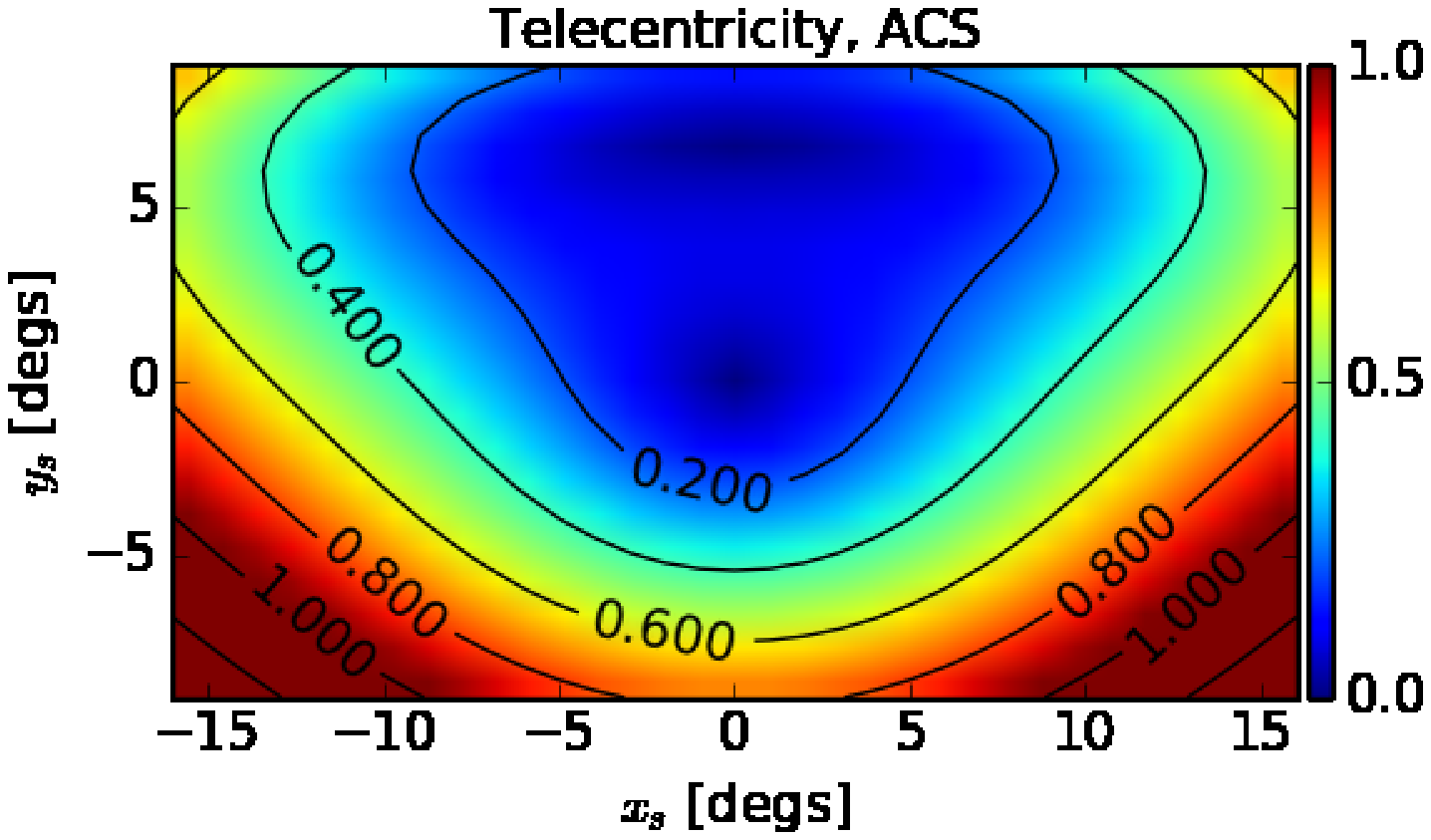}
    \includegraphics[width=0.8\hsize]{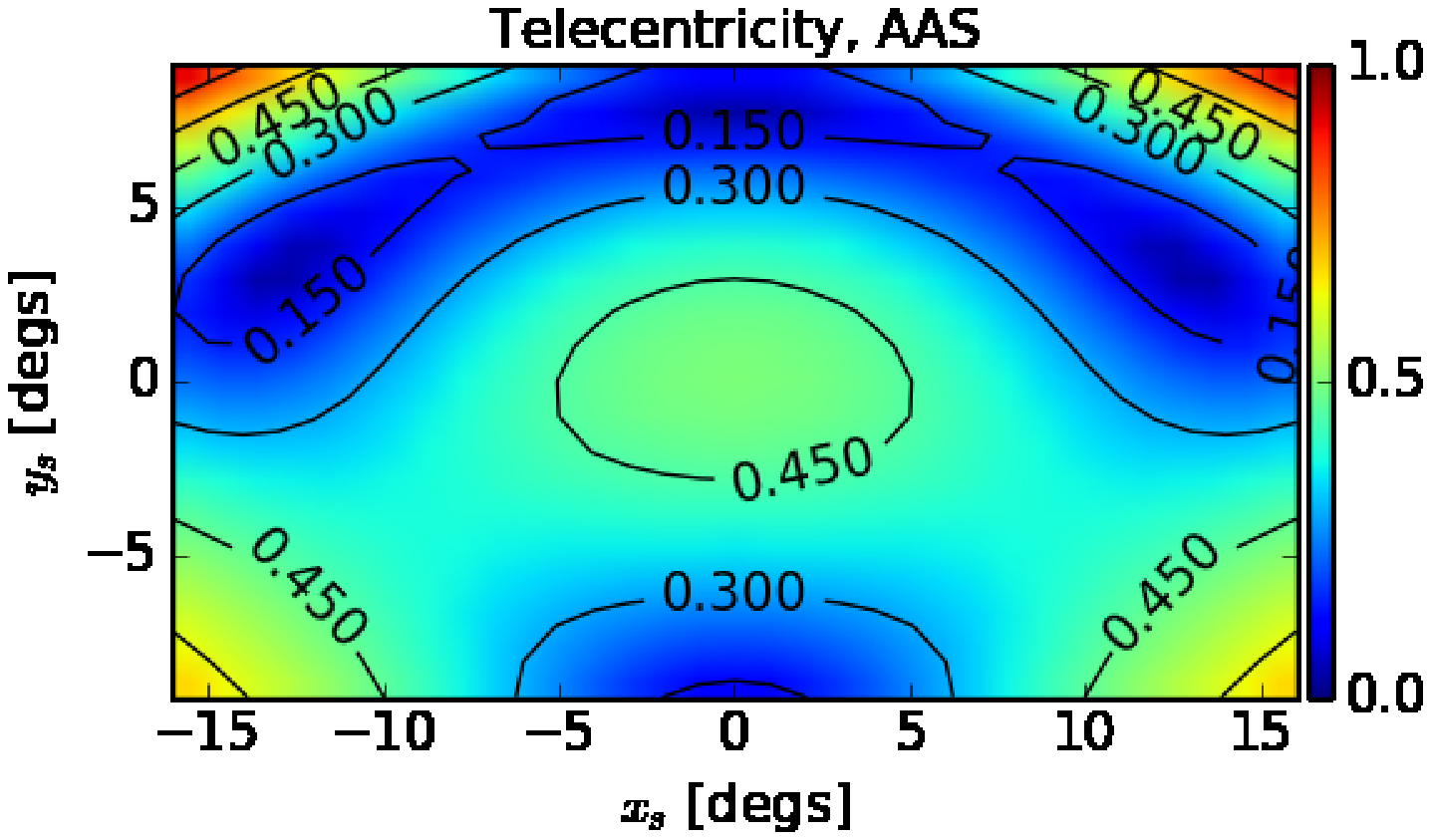}
   \caption{The telecentricity is plotted over the FOV in unit of degree. Top: SCS design. Middle: ACS design. Bottom: AAS design.}
   \label{fig:Tel}
\end{figure}

\subsection{Strehl ratio}

Figure~\ref{fig:SR} shows two-dimensional maps of SR for the square degree FOV with 1 square degree resolution. 
The available focal plane area, defined as SR $>$ 0.95 at 150~GHz for a convenience, is $24\times14$, $28\times16$, $32\times18$ square degrees for SCS, ACS, and AAS cases, respectively. Even if we use conic surfaces only a CD telescope provides a quite large FOV. Moreover, the progressive improvements of SR are obtained in the ACS surface and further particularly in the AAS one over the wide FOV, compared with the SCS one.

The Strehl ratio is high enough, compared to 0.8, if we focus on the outside of 32 × 18 deg$^2$ for the AAS case. However, we have to couple the incident light to detectors. The Strehl ratio of about 0.95 is required in order to keep the coupling higher. Another point is that if we explore a larger FOV, the aperture projection onto the plane normal to the chief ray goes to an ellipse. It will give rise to an elliptic PSF. Therefore, we demonstrated the case of 32 × 18 deg$^2$ FOV.

\subsection{Telecentricity}
Figure~\ref{fig:Tel} shows the two-dimensional map of the incident angle of chief rays into the focal plane for the SCS, ACS, and AAS designs. All the designs satisfy the requirement on the telecentricity, smaller than 1 degree.

\subsection{Variation of $F$ number}
Figure~\ref{fig:Fnum} shows the two-dimensional $F$ number distribution of the AAS.  With the designed $F$ of 3.05 for the AAS, our restriction corresponds to a range of $F$ numbers between 2.96 and 3.14.

\subsection{Third flattening}
Figure~\ref{fig:PSF} shows the third flattening and the rotational directions of the nine PSFs over one side of the FOV. The extracted third flattening and the orientation, $\theta$, is defined in Equation~\ref{eq:psfdef}.

\subsection{Polarization properties}
Although we did not include the polarization properties as a part of the figure-of-merit, it is important to assess its performance. We compute the rotation of the plane of polarization, cross-polarization response of the PSF with respect to the co-polarization, and instrumentally-induced polarization.

\subsubsection{Rotation}
\label{subsec:rotation}
\begin{figure}[!t]
   \centering
     \includegraphics[width=\hsize]{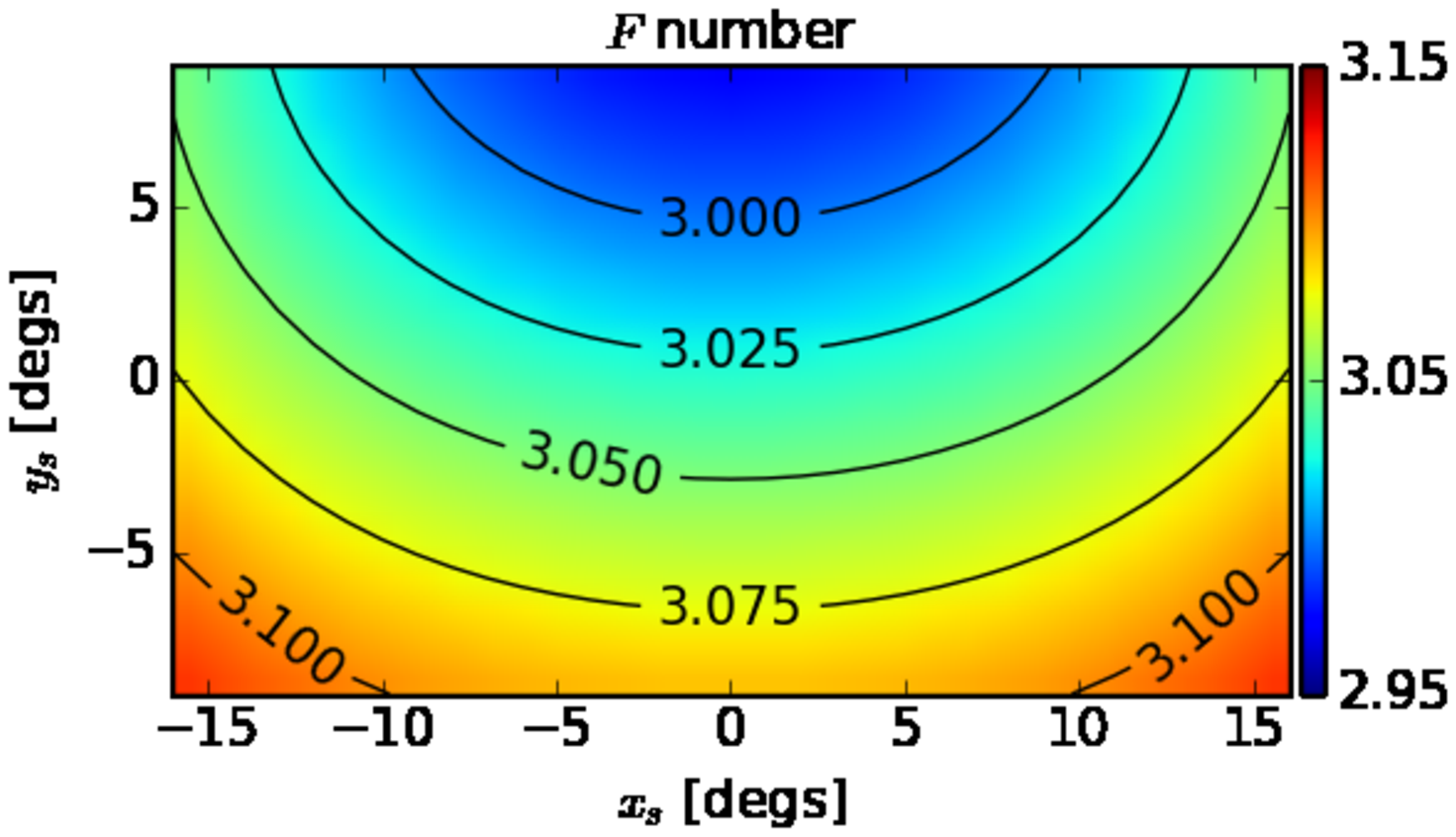} 
   \caption{The $F$ number at each field position for the AAS design.}
   \label{fig:Fnum}
%
   \centering
    \includegraphics[width=\hsize]{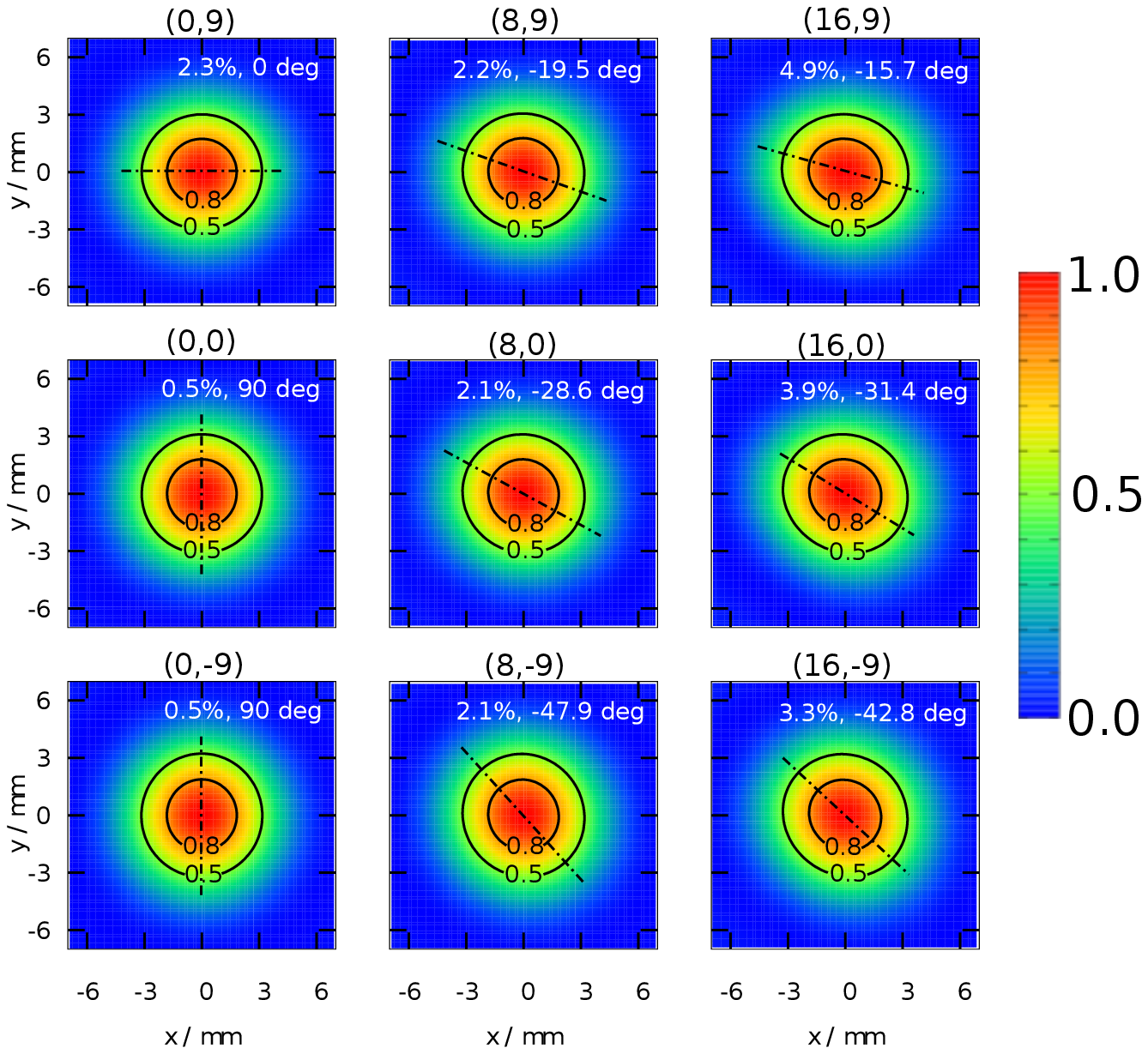} 
   \caption{The PSF at each field position for AAS case. The labels above the panels show the incident angles along $x$ and $y$ axes in degree. The dashed-dotted line represents a major axis. The origin of each panel is set to the position of each peak intensity. The percentage in white shows the third flattening.}
   \label{fig:PSF}
   \centering
    \includegraphics[width=3.4in]{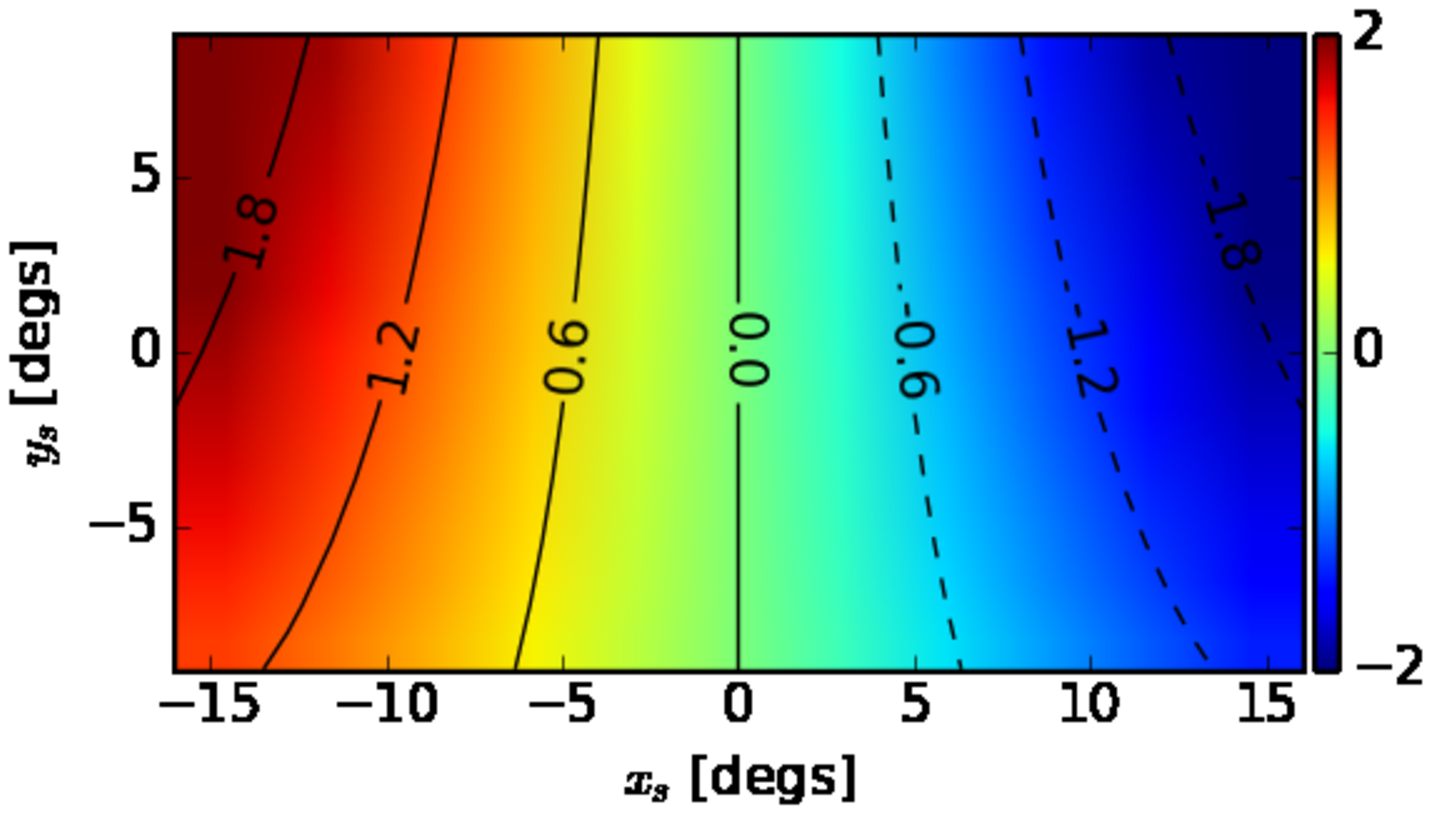}
    \caption{The rotational angle of polarization over the FOV for AAS design.}
   \label{fig:polangle}
\end{figure}

Figure~\ref{fig:polangle} shows the rotational angle of the plane of polarization associated with the chief ray at each field position for AAS design. The results for SCS and ACS designs are nearly identical due to the no significant difference of the incident angle with respect to the primary and secondary mirrors among the surface types.

In this analysis, the angle is derived only from the chief ray, and thus the effect of rotation from the geometrical averaging over the illumination pattern on a mirror surface is not considered. In terms of materials, we account for the finite conductivity of an aluminum as well as an SiO layer, a protection coating on aluminum mirrors. The complex index that is used for this analysis is summarized in Table~\ref{tab:index}. As shown in Table~\ref{tab:index}, the loss tangent of SiO is so small and the thickness is also so thin with respect to wavelength that the effect of an SiO layer on polarization is negligible. The use of this type of a telescope is expected to be used at the cryogenic temperature for a next generation CMB polarization experiment. Thus, we employ the complex index of the mirrors from the data at the available lower temperature, 110~K \cite{diez2000}. On the other hand, the index of the SiO coating is only available for a room temperature value \cite{ordal1988, desai1984, lamb1996, bock1995}. The variation of the rotation over the FOV is less than 2 degrees and this is a part of the quantities to be calibrated.

\subsubsection{Cross polarization} \label{sec:cx-pol}
\begin{figure}[h]
   \centering
    \includegraphics[width=3.8in]{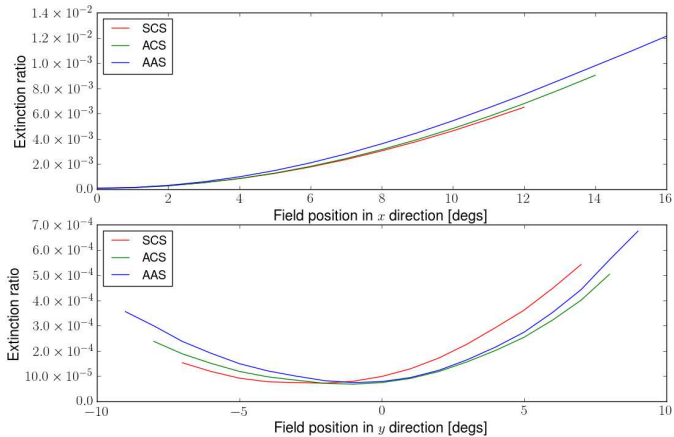}
    \caption{The extinction rate is computed along the $x_s$ and $y_s$ axes for three different surface types. The top panel is evaluated at $y_s = 0$, and the bottom is at $x_s=0$.}
   \label{fig:crosspol}
\end{figure}

Figure~\ref{fig:crosspol} shows the extinction ratio as a measure of cross polarization. We compute the PSF with linearly polarized incident light, and compute the ratio of the peak intensity of the cross response PSF to that of the co-response PSF. 

This ratio is about $10^{-5}$ at the center of the focal plane and the ratio increases as it approaches to the edge of the focal plane. There is no clear difference among the three surface types due to the similarity of the incident angle with respect to the mirror surface at the primary and the secondary mirrors. 

The result depends on the surface shape as well as the complex index of the mirror surface. When we vary the complex index of aluminum by 10\%, the ratio at the center of the focal plane, $10^{-5}$, only changes by at the level of $10^{-10}$. Thus, the choice of the index is not highly sensitive to the results.

\subsubsection{Instrumental polarization} \label{sec:ip}
\begin{figure}[t]
   \centering
\includegraphics[width=0.9\hsize]{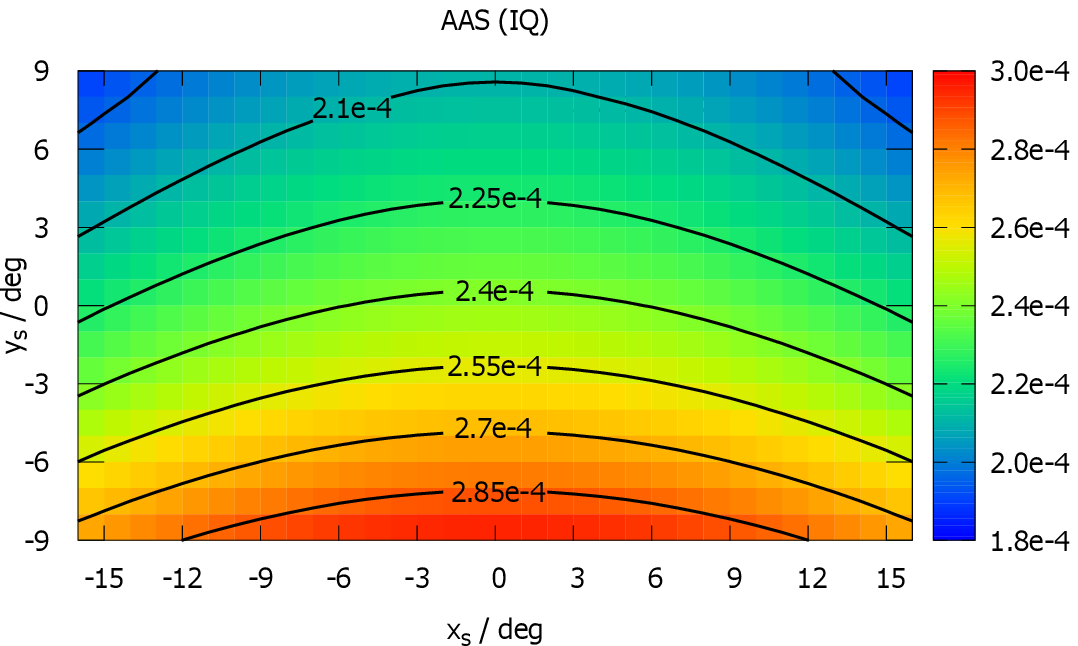}
\includegraphics[width=0.9\hsize]{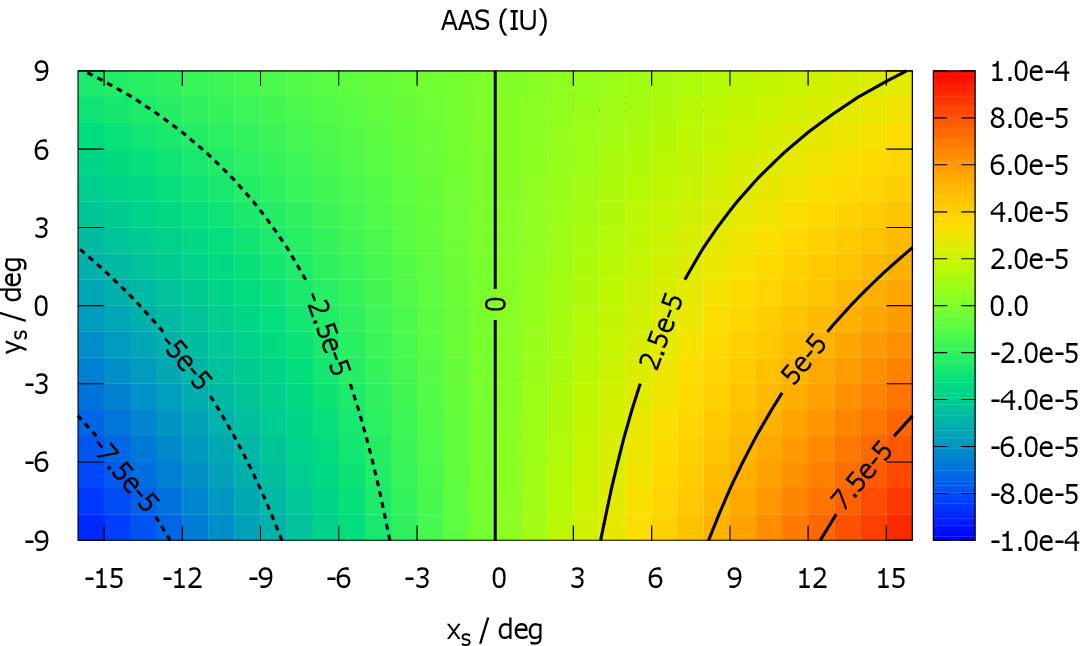}
\includegraphics[width=0.9\hsize]{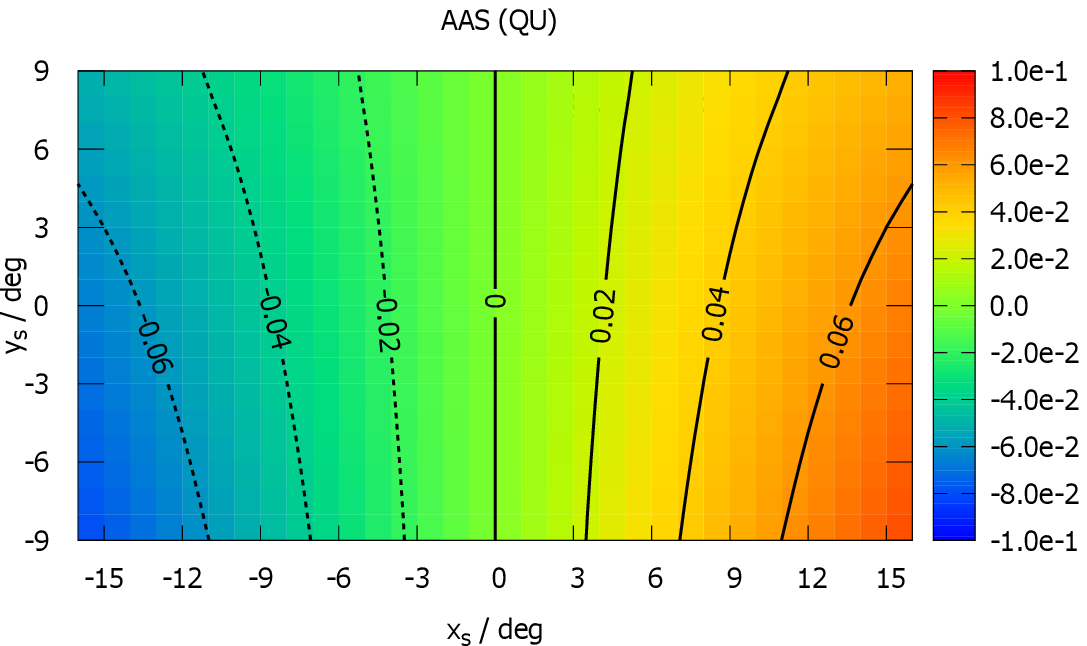}
    \caption{The three elements, $M_{IQ}$, $M_{IU}$, $M_{QU}$, of Mueller matrix over the FOV for the surface type of AAS.}
   \label{fig:mueller}
\end{figure}

Figure~\ref{fig:mueller} shows the three elements, $M_{IQ}$, $M_{IU}$, $M_{QU}$ of Mueller matrix for AAS design. The Mueller matrix is defined as an end-to-end matrix that includes the primary and the secondary surfaces. The matrix element is written as $M_{ij}$, where $i, j = I, Q, U, V$.

This instrumental polarization is produced due to the finite conductivity of the mirror material, and thus it depends on the
\clearpage

\begin{table}[!t]
   \centering
   \caption{The complex index of refraction for the mirror surface.}
   \label{tab:index}
   \begin{tabular}{c|c|c|c} 
   Material & Complex index & Thickness  & Temp. \\ \hline
	Al~\cite{diez2000} & $2543.4 + 2552.2 \ i$ & 1~$\mu$m & 110~K  \\
	SiO~\cite{ordal1988, desai1984, lamb1996, bock1995} & $2 + 8\times10^{-4} \ i$ & 25~nm & room \\ \hline
   \end{tabular}
\end{table}

\hspace{-3.5mm}complex index which we described in Section~\ref{results}.\ref{subsec:rotation}. The instrumental polarization is defined as $I_p = \sqrt{M_{IQ}^2+M_{IU}^2}$. $M_{QU}$ does not contribute the instrumental polarization because $M_{QU}$ has the effect of rotating the polarization angle only. $I_p$ is dominated by $M_{IQ}$ and the highest value in the FOV is about $3\times10^{-4}$ in the negative part of $y$. This is due to the fact that the rays in this position of FOV are reflected at the mirror surfaces steeper angle than the other areas of the FOV. The results for the SCS and the ACS are similar to the AAS case, because these have only minor differences in the surface angles with respect to the ray incident to each mirror surface.

\section{Discussions}
In this section, the obtained designs are discussed from the point of view of CMB polarization observation. There are a lot of figures-of-merit but we pick out a focal plane area, spectral frequency dependence of Strehl ratio, and Mueller matrix to focus on. Finally, we touch on a future prospect.

We define the effective available focal plane area, $A_{fp}$, as the area with the SR above 0.95 at 150~GHz within the $32\times18$ square degrees. 
We place this restriction to the computed area, not based purely on the optical quality, but from practical limits, e.g. the physical interferences among the mirrors, the aperture, and the focal plane.
Even within this bounded FOV, the ratio of this effective focal plane area is $A_{fp,ACS}/A_{fp,SCS}=1.16$ and $A_{fp,AAS}/A_{fp,SCS}=1.25$, respectively.
If one places more detectors at a given band by employing AAS, the ratio directly relates to the gain of the number of detectors. 
Thus, in a single-frequency observation limit, this area ratio is directly related to a gain of a square-root of a mapping speed~\cite{griffin2002}. 

\begin{figure}[t]
   \centering
 \includegraphics[width=3.4in]{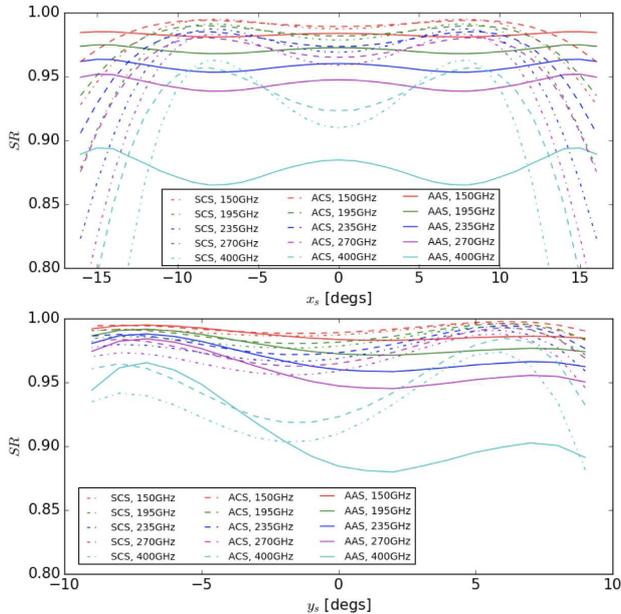}
    \caption{The $SR$ computed at 150~GHz is extrapolated to 195, 235, 270 and 400~GHz using Equation~\ref{eq:SR}. The top panel is the SR sliced along $x_s$ axis with $y_s=0$. The bottom panel is the SR sliced along $y_s$ axis with $x_s=0$.}
   \label{fig:SR_nu}
\end{figure}

All the results shown in this paper are evaluated at the reference frequency $\nu_\mathrm{ref}=150$\,GHz. We discuss the prospective performance at other frequencies. We compute SR based on the RMS wave front aberration $w$, which is independent of the wavelength. Thus, we can extrapolate the computed SR to other frequencies:
\begin{eqnarray}
\label{eq:SR}
SR \simeq e^{ - \left( 2 \pi w \nu / c \right)^2 } = { SR_{ 150 } }^{ \nu^2 / { \nu_\mathrm{ ref } }^2 }
\end{eqnarray}
where $ c $ is the speed of light and $ SR_{ 150 } $ is the SR at 150~GHz.
Based on this relationship, the cuts of the projected diffraction-limited focal plane area are plotted in Figure~\ref{fig:SR_nu}.
The bottom line of this plot shows that the entire FOV is diffraction-limited below 400~GHz.
This allows a full freedom to choose the placement of the detector at any frequency below 400~GHz. 
Therefore, the focal plane design has no constraint of placing a higher frequency detector at the center of the focal plane and the lower frequency detector at the edge of the focal plane due to the limited optical performance.
Implementing beyond 400~GHz is an option by placing them at the central region of the focal plane. 
For higher frequency use, the next limitation may come from the surface accuracy, roughness, and misalignment of the primary and secondary mirrors.

The element of the computed Mueller matrix also depends on the frequency due to the frequency dependence of the complex index. Although we have no data to infer the effect to the other frequencies, we do not expect 10~\% variation of the complex index within a millimeter wave frequency range due to the nature of the SiO and aluminum. Thus, the level of the effect that we estimate still holds within the millimeter wave. In order to judge whether or not this magnitude of the Mueller matrix elements becomes an issue, further study such as the simulation from a telescope to a power spectrum is needed. When they are harmful there are a few methods to mitigate this instrumental polarization effect. One is to use a polarization modulator composed of a half wave plate at the aperture \cite{kusaka2014}.

There seems no significant defects of this telecentric wide FOV telescope relating to optical performances within our study in this paper, therefore, our designs are a candidate for a telescope design for CMB polarization observation. Some performances which are not investigated in this paper will be issues to be addressed, e.g. the size of two large mirrors, stray light, and physical optics performance.

As a further study, although we have limited our discussion with the $F$ number of 3.0 and the entrance pupil diameter of 400~mm in this paper, this optical design can be extended to other $F$ number and the entrance pupil diameter when applied to a ground-based telescope. The design example with using SCS can be found in Bernacki et al.~\cite{bernacki2012}.

In the actual CMB telescope, the study to avoid stray light is needed. It is in progress but one example of a baffling structure is shown in Fig. \ref{fig:layout}. In addition, the coupling between the PSF from the telescope and the beam pattern from the feed-horn is important to understand. This requires the physical optics simulation, and this is also in progress. They are beyond the scope of this paper.

A large diameter radio telescope traditionally employs an on-axis telescope with a single feed-horn. Given increasing interest for a larger FOV in the radio community, the telecentric wide FOV telescope obtained here can be a candidate with two large mirrors instead of the one large primary and the small secondary seen in a conventional radio telescope.

\section{Conclusions}
We report the wide FOV CD telescope using the anamorphic aspherical surfaces with the correction terms up to 10th order. 
We achieve $32\times18$ square degrees with $SR>0.95$ for AAS at 150~GHz. 
This AAS system with this FOV achieves sufficient telecentricity in the image space.
This FOV is diffraction-limited below 400~GHz.
Thus, when one designs a multi-frequency focal plane, one can choose any place to put detectors as long as the detection band of each detector is below 400~GHz. 
Such this design is particularly attractive for use in a multi-color large focal plane CMB telescope from a balloon and space platform.

\section*{Acknowledgments}
The authors appreciate the helpful comments from the reviewers. We would especially like to thank Kimihiro Kimura, Masanori Inoue and Makoto Ito (Osaka Prefecture Univ.), and Hirokazu Ishino (Okayama Univ.) for discussions. 
This work was supported 
by the ISAS/JAXA Phase-A1 program,
by MEXT KAKENHI Grant Number JP15H05891,
by World Premier International Research Center Initiative (WPI), MEXT,
and by the JSPS Core-to-Core program. 

\bibliography{ref_kashima2017}

\begin{thebibliography}{10}

\bibitem{planck2015}
R.~Adam et~al.,
\newblock Astronomy \& Astrophysics {\bf 594}, A1 (2016).

\bibitem{keckbicep2VI}
J.~Bock et~al.,
\newblock arXiv  (2015).

\bibitem{tanaka1975}
H.~{Tanaka} and M.~{Mizusawa},
\newblock Electronics Communications of Japan {\bf 58}, 71 (1975).

\bibitem{dragone1978}
C.~Dragone,
\newblock Bell System Technical Journal {\bf 57}, 2663 (1978).

\bibitem{tran2008}
H.~Tran, A.~Lee, S.~Hanany, M.~Milligan, and T.~Renbarger,
\newblock Applied optics {\bf 47}, 103 (2008).

\bibitem{hanany2013}
S.~Hanany, M.~D. Niemack, and L.~Page,
\newblock Cmb telescopes and optical systems,
\newblock in {\em Planets, Stars and Stellar Systems}, pages 431--480,
  Springer, 2013.

\bibitem{imbriale2011}
W.~A. Imbriale, J.~Gundersen, and K.~L. Thompson,
\newblock IEEE Transactions on Antennas and Propagation {\bf 59}, 1972 (2011).

\bibitem{essinger2011}
T.~Essinger-Hileman,
\newblock {\em Probing inflationary cosmology: The Atacama B-mode search
  (ABS)},
\newblock PhD thesis, Princeton University, 2011.

\bibitem{tauber2014}
J.~A. Tauber, P.~Nielsen, and A.~Martin-Polegre,
\newblock The planck mission and its optical system,
\newblock in {\em Antennas and Propagation (EuCAP), 2014 8th European
  Conference on}, pages 2588--2590, IEEE, 2014.

\bibitem{tran2010}
H.~Tran et~al.,
\newblock Optical design of the epic-im crossed dragone telescope,
\newblock in {\em SPIE Space Telescopes and Instrumentation}, Society of
  Photo-Optical Instrumentation Engineers, 2010.

\bibitem{sugai2016}
H.~Sugai et~al.,
\newblock Optical designing of litebird,
\newblock in {\em SPIE Astronomical Telescopes+ Instrumentation}, pages
  99044H--99044H, International Society for Optics and Photonics, 2016.

\bibitem{matsumura2016}
T.~Matsumura et~al.,
\newblock Journal of Low Temperature Physics {\bf 184}, 824 (2016).

\bibitem{niemack2016}
M.~D. Niemack,
\newblock Applied optics {\bf 55}, 1688 (2016).

\bibitem{arxiv1705.02170}
P.~de~Bernardis~et al. and for~the CORE~collaboration,
\newblock arxiv.1705.02170  (2017).

\bibitem{ishino2016}
H.~Ishino et~al.,
\newblock Litebird: lite satellite for the study of b-mode polarization and
  inflation from cosmic microwave background radiation detection,
\newblock in {\em SPIE Astronomical Telescopes+ Instrumentation}, pages
  99040X--99040X, International Society for Optics and Photonics, 2016.

\bibitem{CodeV}
{Synopsys, Inc.},
\newblock \textit{CODE V Electronic Document Library}, 2015.

\bibitem{mahajan1983}
V.~N. Mahajan,
\newblock Journal of the Optical Society of America (1917-1983) {\bf 73}, 860
  (1983).

\bibitem{gerrard2012}
A.~Gerrard and J.~M. Burch,
\newblock {\em Introduction to matrix methods in optics},
\newblock Courier Corporation, 2012.

\bibitem{buchdahl1993}
H.~A. Buchdahl,
\newblock {\em An introduction to Hamiltonian optics},
\newblock Courier Corporation, 1993.

\bibitem{diez2000}
M.~C. Diez, T.~O. Klaassen, K.~Smorenburg, V.~Kirschner, and K.~J. Wildeman,
\newblock Reflectance measurements on submillimeter absorbing coatings for
  hifi,
\newblock in {\em Astronomical Telescopes and Instrumentation}, pages 129--139,
  International Society for Optics and Photonics, 2000.

\bibitem{ordal1988}
M.~A. Ordal, R.~J. Bell, R.~W. Alexander, L.~A. Newquist, and M.~R. Querry,
\newblock Applied optics {\bf 27}, 1203 (1988).

\bibitem{desai1984}
P.~D. Desai, H.~James, and C.~Y. Ho,
\newblock Journal of physical and chemical reference data {\bf 13}, 1131
  (1984).

\bibitem{lamb1996}
J.~W. Lamb,
\newblock International Journal of Infrared and Millimeter Waves {\bf 17}, 1997
  (1996).

\bibitem{bock1995}
J.~Bock, M.~Parikh, M.~Fischer, and A.~Lange,
\newblock Applied optics {\bf 34}, 4812 (1995).

\bibitem{griffin2002}
M.~J. Griffin, J.~J. Bock, and W.~K. Gear,
\newblock Applied Optics {\bf 41}, 6543 (2002).

\bibitem{kusaka2014}
A.~Kusaka et~al.,
\newblock Review of Scientific Instruments {\bf 85}, 024501 (2014).

\bibitem{bernacki2012}
B.~E. Bernacki et~al.,
\newblock Wide-field-of-view millimeter-wave telescope design with ultra-low
  cross-polarization,
\newblock in {\em SPIE Defense, Security, and Sensing}, pages 836207--836207,
  International Society for Optics and Photonics, 2012.

\end{thebibliography}

\end{document}